\documentclass[
 amsmath,amssymb,
 aps,
reprint,%
author-numerical,%
superscriptaddress,
]{revtex4-2}
\usepackage{mathrsfs}
\usepackage{graphicx}% Include figure files
\usepackage{dcolumn}% Align table columns on decimal point
\usepackage{bm}% bold math

\preprint{AAPM/123-QED}

\begin{document}

\title{Effects of the antiferrodistortive instability on the structural behavior of BaZrO$_3$ \\ by atomistic simulations}% Force line breaks with \\

\author{M. Sepliarsky\thanks{corresponding author}}
\email{sepliarsky\@ifir-conicet.gov.ar}
\affiliation{Instituto de Física Rosario (CONICET) and Facultad de Ciencias Exactas, Ingenier\'\i{}a y Agrimensura,Universidad Nacional de Rosario, Rosario, Argentina}
\author{R. Machado}%
\affiliation{Instituto de Física Rosario (CONICET) and Facultad de Ciencias Exactas, Ingenier\'\i{}a y Agrimensura,Universidad Nacional de Rosario, Rosario, Argentina}
\author{S. Tinte}
\affiliation{Instituto de Física del Litoral-CONICET and Facultad de Ingeniería Química, Universidad Nacional del Litoral, Santa Fe, Argentina}
\author{M. G. Stachiotti}
\affiliation{Instituto de Física Rosario (CONICET) and Facultad de Ciencias Exactas, Ingenier\'\i{}a y Agrimensura,Universidad Nacional de Rosario, Rosario, Argentina}

\date{\today}% It is always \today, today,
             %  but any date may be explicitly specified

\begin{abstract}
Recently, the possibility of a low-temperature non-cubic phase in BaZrO$_3$ has generated engaging discussions about its true ground state and the consequences on its physical properties. In this paper, we 
investigate the microscopic behavior of the BaZrO$_3$ cubic phase by developing a shell model from $ab~initio$ calculations and by performing molecular dynamics simulations at finite temperature and under negative pressure.
We study three different scenarios created by tuning the intensities of the antiferrodistortive (AFD) instabilities, and consequently, the sequence of phase transitions with temperature.
From a detailed analysis of the cubic phase at atomic scale, we find that oxygen octahedra are barely distorted, 

present rotation angles that may oscillate with significant amplitudes, are AFD
correlated with their closest neighbors on the plane perpendicular to the pseudocubic rotation axis exhibiting $(0 0 a^-)$-type ordering,
and form instantaneous, dynamic and unstable domains over time.
Our simulations support the existence of nanoregions with short-range ordering in cubic BaZrO$_3$ associated with experimentally observed anomalies and unveil 
that they can exist regardless of whether or not structural phase transitions related with AFD distortions occur at lower temperatures.

\end{abstract}

\maketitle

\section{Introduction}

ABO$_3$-type oxides constitute an important class of materials with high technological value in numerous device applications. Despite their chemical simplicity, these perovskite compounds are crystallographically subtle, because they exhibit a wide variety of complex structural instabilities, such as ferroelectric, antiferroelectric and antiferrodistortive distortions.
Among them, BaZrO$_3$ based ceramics are of interest in the development of microwave capacitors~\cite{pari_12}, oxide fuel cells~\cite{bae_17,duan_15}, thermal barrier coating~\cite{liu_18}, and as a substrate for high-temperature superconductors~\cite{macma_04,miura_16}. It is also an endpoint of the eco-friendly Ba(Zr,Ti)O$_3$ and (CaBa)(ZrTi)O$_3$ solid solutions, technologically important electroceramic materials due to their high dielectric and piezoelectric constants~\cite{acos_17,liu_09,hao_12,hana_19}. 
From a fundamental point of view, concerns about the existence of low-temperature non-cubic phase have arisen recently.
Indeed, the structural behavior of BaZrO$_3$ at low temperatures has been the subject of an interesting debate with controversial positions. Experimental evidence indicates that BaZrO$_3$ remains in the cubic phase with $Pm\bar{3}m$ symmetry down to zero Kelvin~\cite{dobal_01,akba_05,perri_20,knig_20}, whereas the possibility of a lower-symmetry phase or the existence of regions with short-range ordering arose from theoretical calculations.
Several first-principles studies of the lattice dynamics in BaZrO$_3$ revealed the instability of a non-polar mode at the point $R$ of the Brillouin zone~\cite{zhong_95,akba_05,ben_06,bilic_09,lebe_13}. The $R_{25}$ symmetry mode corresponds to a rigid antiphase rotation of the oxygen octahedra that would give rise to a ground state of symmetry that condenses in antiferrodistortive (AFD) structures, similar to that observed in SrTiO$_3$~\cite{shir_69}. Nevertheless, the instability corresponding to these AFD displacements are rather weak, and results are very sensitive to the choice of the exchange-correlation functional. In fact, the instability of the cubic phase obtained with LDA decreases with PBSol and disappears with the use of hybrid functionals~\cite{perri_20,gran_20}. Experimentally, the possibility of a low-symmetry structure is associated with the observation of enhanced values in the Debye-Waller factor for the Ba-O pair and disagreements
between experimental and theoretical low-temperature dielectric constant~\cite{lebe_13,akba_05,ben_06}. Recently, a local structure study combining electron diffraction and total neutron scattering reported a structural change below 80~K that authors related to the presence of correlated octahedral rotations in regions of about 3~nm in size, although the static or dynamic nature of these regions could not be elucidated~\cite{levin_21}.

Despite the progress made by using the state of the art experimental techniques and the most precise theoretical methods, the controversies about the low-temperature behavior of BaZrO$_3$ are not completely clarified, and additional studies are required. In this sense, simulations with atomic-level models are a powerful tool to provide microscopic insights into systems where a large number of atoms are required and/or finite-temperature effects have to be taken into account. Among the different approaches used to describe perovskite oxides, effective hamiltonians~\cite{akba_05,zhan_21} and bond valence models~\cite{raite_11,zhang_22} were presented to study BaZrO$_3$, however, the nature of its ground state was not addressed in detail in such works. Here we develop a shell model potential for BaZrO$_3$ to provide insights into 
the relationship that exists between the anomalies observed in the cubic phase and the possibility of an effective phase transition or the appearance of a short-range ordering of low symmetry.
By performing molecular dynamics simulations, we conclude that the experimentally observed indications in BaZrO$_3$ related to a possible low-temperature phase transition can be found in its cubic phase regardless of whether such a transition exists or not.

The work is organized as follows. In Section~II, the procedure to develop the model and validate the new BaZrO$_3$ potential is introduced. Then
we analyze the macroscopic (Sec.~III) and atomic-scale (Sec.~IV) behavior of the compound under three different scenarios obtained by varying the AFD instability strength. Finally, conclusions are presented in Section~V.

%**************************************************************************************
\section{Model development}
%*************************************************************************************

Simulations using the shell model have been extensively employed to study properties of perovskites oxides. In this model, each atom is described by two charged particles: a core connected to a massless shell, which accounts phenomenological for atomic polarizability. This classical model takes into account electrostatic interactions among cores and shells of different atoms, and short-range interactions between shells. The data used to fit the model parameters are obtained from first-principle results of key properties of the compound, such as total energies, stress components, atomic forces, phonon frequencies, eigenvectors, etc.

\subsection{Potential fitting}

Here, we develop a new shell model potential for BaZrO$_3$ that contains isotropic fourth-order core-shell couplings ($k_2$, $k_4$), long-range Coulombic interactions and short-range interactions described by two different types of potentials. A Born-Mayer potential 
$V(r)~=~A\exp(-r/\rho)$ is used for the Ba-O and Zr-O pairs, and a Buckingham potential $V(r)~=~A\exp(-r/\rho)+C/r^{-6}$ is used for O-O interactions. Rather than fitting the complete set of parameters for BaZrO$_3$, we take advantage of our previously developed BaTiO$_3$ model. BaTiO$_3$ and BaZrO$_3$ share the same perovskite structure ABO$_3$ and differ only in the element at the B-site. Similarly, we consider that the model potential for BaTiO$_3$ and BaZrO$_3$ differs only in B-cation related interactions while those present in both compounds (O-O and Ba-O) are the same. Then, we transfer from the BaTiO$_3$ model the coefficients of similar interactions and fit only the new parameters of the potentials involving Zr atoms, namely, polarizability parameters, core, and shell charge distribution, and the coefficients of the Zr-O short-range Born-Mayer interaction. The approximation considers that the interaction between atoms does not strongly depend on their local environment and is still valid in situations different from those contemplated during the adjustment. This assumption has already been used to study various compounds within the shell model~\cite{tinte04,sepli11}, and 
the extension to BaTiO$_3$ and BaZrO$_3$ also rely on first-principles results~\cite{amor_18}. 

Despite the composition similarity, both compounds display very different properties. BaTiO$_3$ is a well-studied ferroelectric that undergoes successive structural phase transitions going from cubic to tetragonal to orthorhombic and rhombohedral phases as temperature decreases. All these phases involve small displacements of the cations with respect to the oxygen octahedra along $[001]$, $[011]$ and $[111]$ directions respectively~\cite{kwei_93}. According to first-principles calculations, the change in properties when Ti is replaced by Zr can mainly be attributed to differences in the cation-oxygen interactions while other force constants do not exhibit relevant variations from one compound to the other~\cite{amor_18}.

The BaTiO$_3$ model used a starting point accurately  
reproduces the temperature sequence of phase transitions of the compound and it was used to investigate intrinsic effects of Mg impurities on BaTiO$_3$ ferroelectric properties~\cite{macha_19}. Following the procedure described in that work, input data to adjust the new parameters are obtained from ab-initio results using the general gradient approximation for solids (PBEsol)~\cite{pbesol} as implemented in the VASP package~\cite{vasp}. Calculations are performed considering a 40-atom supercell, which allows one to take into account low-symmetry structures compatible   the condensation of the rotational $R_{25}$ mode. The database is made up of values of total energies, forces on atoms, and stresses of a total of 28 different configurations, including those relevant with cubic $Pm\bar{3}m$, tetragonal $I4/mcm$, orthorombic $Imma$, and rhombohedral $R\bar{3}c$ symmetry. It also contains configurations with displaced Zr atoms and structures under strain. The model parameters are adjusted using a least-square method. The complete set of parameters of the resulting potentials, including those of Ti-related interactions, are listed in Table \ref{poten}.

\begin{table}
\caption{\label{poten}%
Shell model parameters for BaZrO$_3$. Ti-related parameters are also included for completeness. For all parameters the units of energy, length, and charge are given in eV, \AA \ and electrons respectively.}
\begin{ruledtabular}
\begin{tabular}{ccccc}
Atom & \textrm{Core charge} & \textrm{Shell charge} & \textrm{$k_2$} & \textrm{$k_4$} \\
\hline
Ba & 3.93 & -2.02 & 193.2 & 8331.7 \\
Ti & 4.39 & -1.14 & 323.1 & 4.3 \\
Zr & 8.79 & -5.54 & 995.5 & 146.3 \\
O  & 0.87 & -2.59 & 31.4  & 6069.5 \\
\hline
Short range & {\it{A}} & {$\rho$} & \it{C} &\\
\hline
Ba-O &  1341.68  &  0.349902  & 0 &\\   
Ti-O &  3794.63  &  0.257287  & 0 &\\   
Zr-O &  4121.62  &  0.272722  & 0 &\\ 
O-O  &   661.20  &  0.316019  &  6.819 & \\
\end{tabular}
\end{ruledtabular}
\end{table}

\begin{table}[b]
\caption{\label{tab:table4}%
Comparison of properties between BaTiO$_3$ and BaZrO$_3$ 
 Values of lattice parameter, and frequencies of the lowest energy transverse optical phonon mode at $\Gamma$-point and $R_{25}$ mode for the cubic structure for BaTiO$_3$ and BaZrO$_3$ obtained with the shell model. Values from ab-initio and experimental results are also included for comparison}
\begin{ruledtabular}
\begin{tabular}{ccrrl}
 &  &   Model & PBEsol & Experiments \\
\hline
$a_{eq}$ & BaTiO$_3$ & 3.980 & 3.984 & 4.003~\footnotemark[1] \\
         & BaZrO$_3$ & 4.199 & 4.192 & 4.1908~\footnotemark[2]\\
\hline
$\Gamma_{TO1}$ & BaTiO$_3$ & -267 & -197 & soft \\
               & BaZrO$_3$ &   91 &   101 &    \\
\hline
$R_{25}$  & BaTiO$_3$ & 171 & 135 &  \\
          & BaZrO$_3$ & -59 & -43 &    \\
      
\end{tabular}
\end{ruledtabular}
\footnotetext[1]{Ref~\cite{kwei_93}}
\footnotetext[2]{Ref~\cite{akba_05}}
\end{table}

\subsection{Validation of transferability of interatomic potentials}

In order to test the capability of the description with transferable interactions to properly describe the behavior observed in BaZrO$_3$, we calculate the following essential properties of BaTiO$_3$ and BaZrO$_3$ in the high-symmetry cubic structure: the equilibrium lattice parameters and the frequencies of the lowest energy transverse optical phonon mode at $\Gamma$-point and the $R_{25}$ mode (see Table 2). The model results are in good quantitative agreement with the reference first-principle results  and experimental data~\cite{akba_05} indicating that the developed potentials adequately reproduce the changes that take place when Zr replaces Ti in the structure composition. The equilibrium lattice parameter of BaZrO$_3$, in particular,  exceeds that of BaTiO$_3$ by more than 5\%, which is due to the size difference between the cations. The larger size of Zr relative to that of Ti manifests in the model through a more repulsive Zr-O short-range interaction.

The absence of ferroelectricity in BaZrO$_3$ is another feature well captured by the model. The ferroelectric behavior in oxide perovskites is usually related to the instability of the lowest energy transverse optical phonon mode at $\Gamma$-point of the cubic structure. The so-called soft mode in BaTiO$_3$ has an imaginary frequency and involves displacements of the Ba and Ti cations against the oxygen anions. Replacement of Ti by Zr hardens that mode and its frequency becomes positive. This mode in BaZrO$_3$, named as last mode, involves mainly Ba displacements against the surrounding oxygen ZrO$_6$ octahedra. The change can be explained by considering ionic-size mismatches according to the Goldschmidt tolerance factor~\cite{gold_26}. Nevertheless, the simple volumetric argument is not enough to explain the transformation from BaTiO$_3$ to BaZrO$_3$, and electronic effects have also to be taken into account. Changes in the Zr-O hybridization mechanism with respect to the Ti-O one also contribute to the absence of polar instabilities in BaZrO$_3$~\cite{amor_18}. This feature is phenomenological considered in the model through a lower polarizability (higher core-shell coupling) of the Zr with respect to that of Ti. 

Finally, the new model also reproduces the softening of the phonon mode associated with cooperative antiphase rotations of the oxygen octahedra. The frequency of this triply degenerate mode at $R$-point in the cubic Brillouin zone decays from 150 cm$^{-1}$ in BaTiO$_3$ to 59$i$~cm$^{-1}$ in BaZrO$_3$,
and from 135 cm$^{-1}$ to 43$i$ cm$^{-1}$ according to {\it ab~initio} calculations using the PBEsol functional~\cite{perri_20}. In the harmonic approximation, the phonon mode depends only on the force constants between oxygen atoms and, therefore, we can consider that O-O interaction plays a predominant role. Note that the parameters of this particular interaction were not fitted or modified for the BaZrO$_3$ model. They come directly from the potential developed for BaTiO$_3$, where AFD distortions were not explicitly considered in the database used to fit the model parameters. The ability of the model to capture the changes in the oxygen octahedra behavior between both compounds is a strong test supporting the validity of the transferability of interatomic interactions and the robustness of the model description. 

\subsection{BaZrO$_3$ behavior at T = 0~K}
\begin{figure}
\centering
\includegraphics[width=\linewidth]{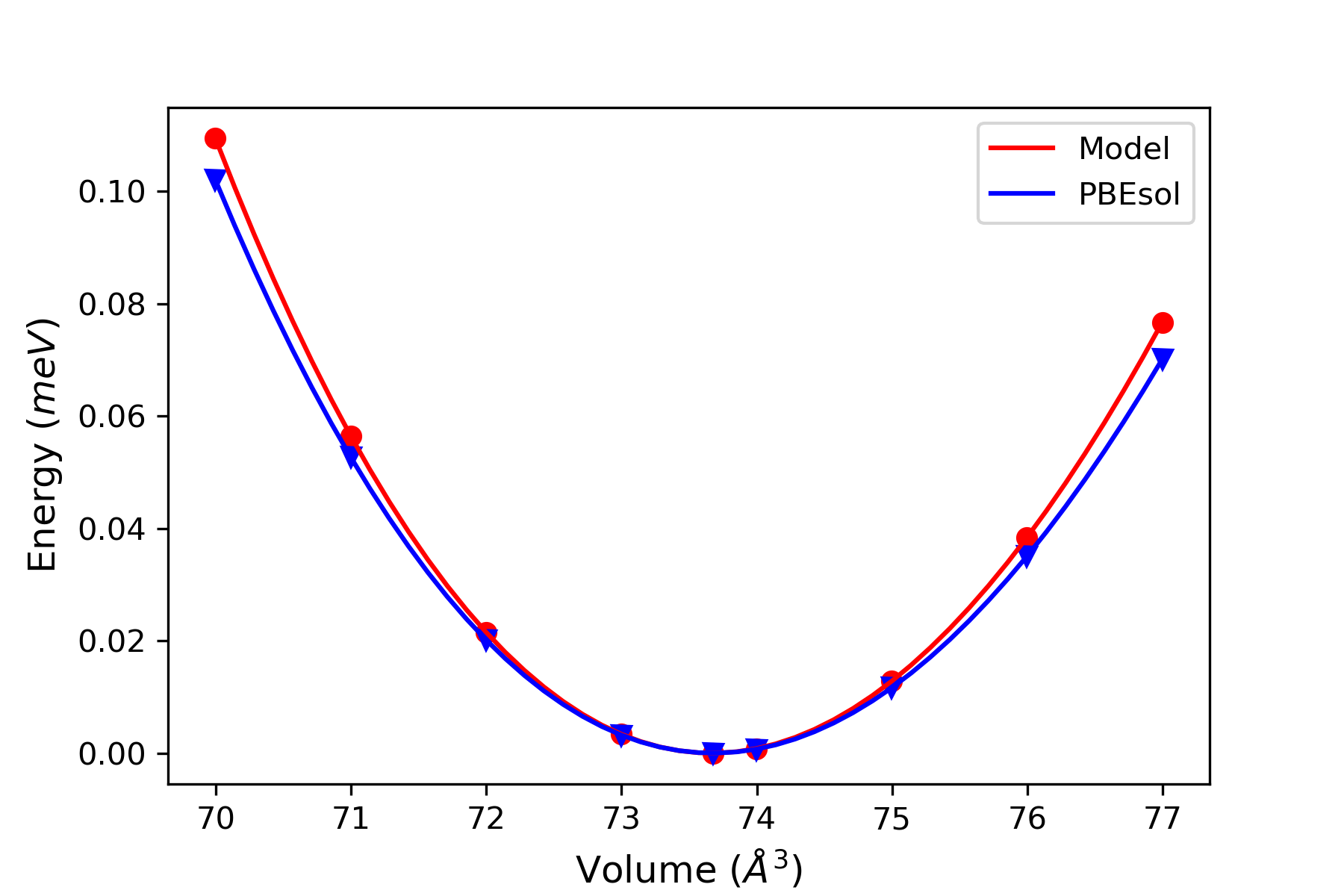}
\caption{\label{fig:Fit}Energy as a function of volume for cubic BaZrO$_3$ from model results and {\it ab~initio} calculations using the PBEsol functional.}
\end{figure}

After validating the developed model, we now focus on the specific behavior of BaZrO$_3$. Figure~\ref{fig:Fit} shows that the energy as a function of the volume obtained with the model agrees well with PBEsol calculations. We obtain a bulk modulus of 175.6~GPa with the model and 162.0~GPa with PBEsol by fitting a Murnaghan equation of state~\cite{murn_44}, both results are close to the experimental value of 189~GPa~\cite{yang_14}.

The unstable $R$-point mode in the high-symmetry structure gives rise to the stabilization of three different configurations with tetragonal ($I4/mcm$), orthorhombic ($Imma$), and rhombohedral ($R\bar{3}c$) symmetry. Specifically, each of them can be obtained by applying AFD distortions about one, two or three of the cubic axes respectively, and then fully relaxing the structures. The energy gain of each structure with respect to that of the cubic configuration listed in Table 3 indicates that the model satisfactorily reproduces the slight stability of these structures, yielding intermediate values between those obtained using the PBEsol and LDA functionals~\cite{lebe_13}. The model, however, predicts a ground state of $R\bar{3}c$ symmetry while the $I4/mcm$ phase is the most stable according to different first-principle calculations. We do not regard this disagreement as significant since in all the cases, the distorted phases differ in energy by less than 1~meV/f.u., therefore the ground state can be considered quasi-degenerate~\cite{amor_18,toul_19}. Furthermore, we are primarily interested in investigating the BaZrO$_3$ behavior near a possible transition from the cubic to a lower-symmetry phase and not another additional structural transitions.

\begin{table}[b]
\caption{\label{tab:table3}%
Comparison of energies for different distorted phases of BaZrO$_3$ obtained with the shell model and ab-initio results. The energy of the cubic $Pm\bar{3}m$ phase is taken as the reference.
}
\begin{ruledtabular}
\begin{tabular}{llccc}
 Phase       & AFD pattern & Model & PBEsol& LDA~\footnotemark[1]   \\
\hline
$I4/mcm$     & $a^0a^0c^-$ & -4.27 & -1.02 & -10.01 \\
$Imma$       & $a^0b^-b^-$ & -4.66 & -0.67 &  -9.47 \\
$R\bar{3}c$ & $a^-a^-a^-$ & -4.79 & -0.37 &  -9.17 \\

\footnotetext[1]{Ref~\cite{lebe_13}}
\end{tabular}
\end{ruledtabular}
\end{table}

\section{Macroscopic properties}
\subsection{Properties at T = 0K}

Having noted that the different distorted phases can be obtained in the model from the condensation of the AFD phonon mode, we now investigate the behavior that BaZrO$_3$ may exhibit in the presence or absence of AFD instabilities. It is well known that AFD distortions are very sensitive to the volume of the system so they can be controlled by applying an external pressure. Specifically, they will be accentuated by applying pressure (lattice compression) or reduced, as we will see here, under negative pressure conditions (lattice expansion).  

\begin{figure}
\centering
\includegraphics[width=\linewidth]{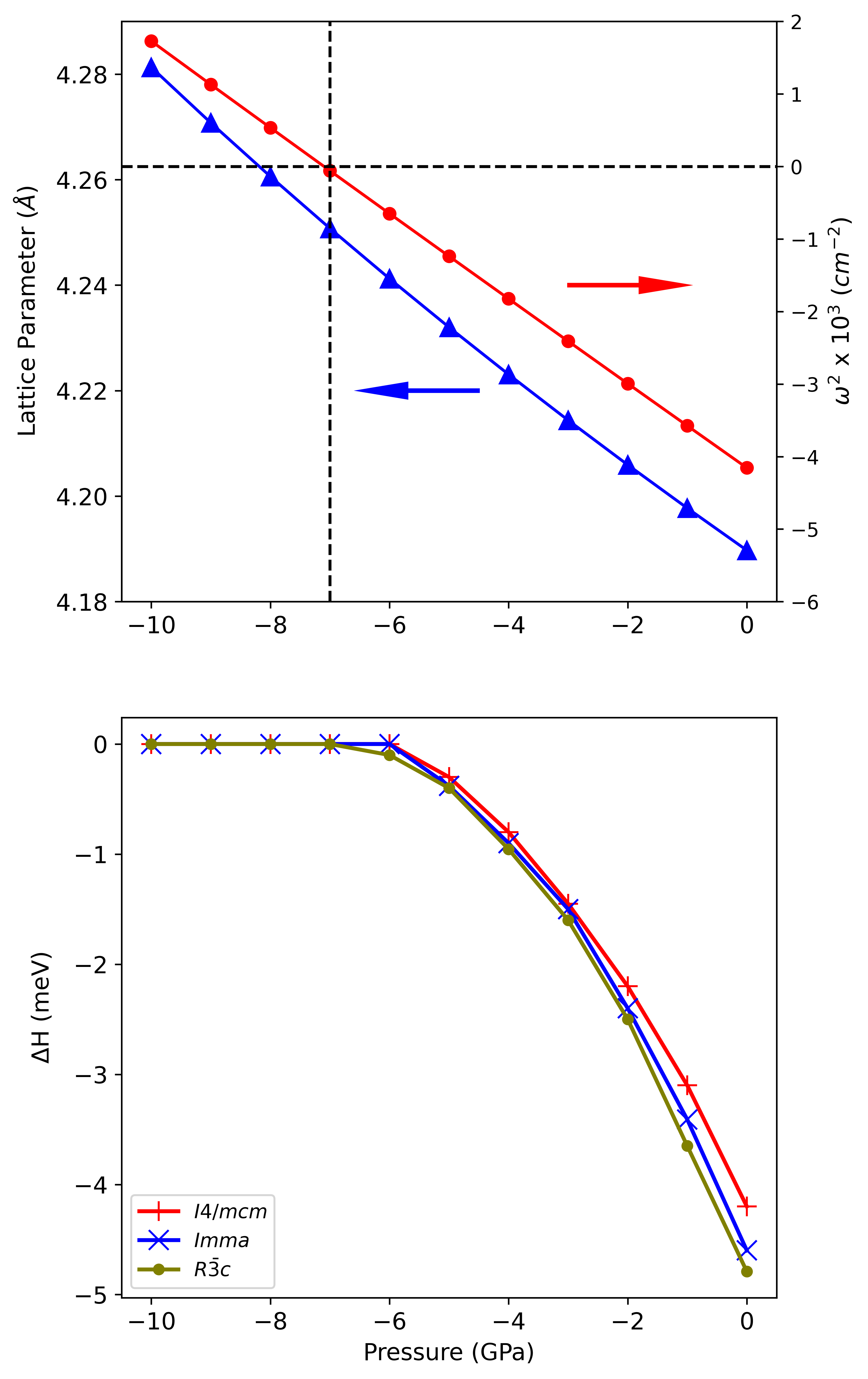}
\caption{\label{fig:LatP}Pressure dependence of the cubic lattice parameter and the square of the $R_{25}$-mode frequency (top panel), and of the enthalpy (bottom panel).}
\end{figure}

Figure~\ref{fig:LatP} displays the evolution of the cubic lattice constant and the square of the $R_{25}$-mode frequency as a function of negative pressure. 
As seen, both magnitudes are quite sensitive to pressure, and the pressure-induced transition to the cubic phase is experienced at $P \approx$ -7~GPa where the frequency becomes positive. At this pressure, the lattice parameter takes a value of 4.251~\AA~which is 1.2~\% greater than that corresponding to $P$ = 0.
Note that this negative pressure 
for the model is higher 
than that of the ab-initio results in Ref.~\cite{gran_20} due to the deeper AFD instabilities and, consequently, our curve is shifted to the left.
In the case of $\omega^2$, the model gives a linear behavior with a slope of -588~cm$^{-2}$/GPa (-9.0~meV$^{2}$/GPa) which in reasonable agreement although slightly higher than the reported value of -7.5~meV$^2$/GPa from ab-initio calculations using a PBE functional~\cite{gran_20}. 

The relative stability of the distorted phases under pressure can be determined by calculating the enthalpy $H = E + PV$, where $E$ and $V$ represent the energy and the volume of the particular structure. The evolution of the enthalpy confirms that the transition to the cubic phase as negative pressure increases takes place at $P=$-7 GPa. Since the difference between the values of the structures is very small, the bottom panel of Figure~\ref{fig:LatP} shows the excess of enthalpy with respect to the cubic phase $\Delta H = H - H_C$ at each pressure value. As seen, the curves of the three low-symmetry phases are practically superimposed, and they continuously approach zero as AFD instabilities vanish. 

%%%%%%%%%%%%%%%%%%%%%%%%%%%%%%%%%%%%%%%%%%%%%%%%%%
\subsection{Finite temperature simulations}
%%%%%%%%%%%%%%%%%%%%%%%%%%%%%%%%%%%%%%%%%%%%%%%%%%
We now apply the developed interatomic potentials for BaZrO$_3$ to simulate the finite temperature behavior of the compound. Molecular dynamics (MD) simulations are carried out using the DL-POLY code~\cite{dlpoly} within a constant stress and temperature (N,$\sigma$,T) ensemble. In this way, the size and shape of the simulation cell are dynamically adjusted in order to obtain the desired average pressure. A supercell of $12\times 12\times12$ 5-atom unit cells (8640 atoms) is used with periodic boundary conditions. The runs are made at temperature intervals of 5~K, and with a time step of 0.4~fs. Each MD run consists of at least 60000 time steps for data collection after 20000 time steps for thermalization. In addition to the lattice parameter, we adopt the antiphase oxygen octahedron tilting as order parameter defined as 
$\theta^{\alpha}=\frac{1}{N}\sum \theta_{i}^{\alpha}\left ( -1 \right )^{n_x(i)+n_y(i)+n_z(i)}$,
where $\theta_{i}^{\alpha}$ corresponds to the octahedra-rotation angle of the $i$-th cell along the rotation axis $\alpha$, and $n_x(i)$, $n_y(i)$, and $n_z(i)$ are integers that label the position of that cell inside the pseudocubic box. We also evaluate the behavior of the polarization but no signs of polar order were observed, indicating that the system remains paraelectric in the range of the simulated temperatures.  

%%%%%%%%%%%%%%%%%%%%%%%%%%%%%%%%%%%%%%%%%%%%%%%%%%%%
\subsection{Temperature-pressure phase diagram}
\begin{figure}
\centering
\includegraphics[width=\linewidth]{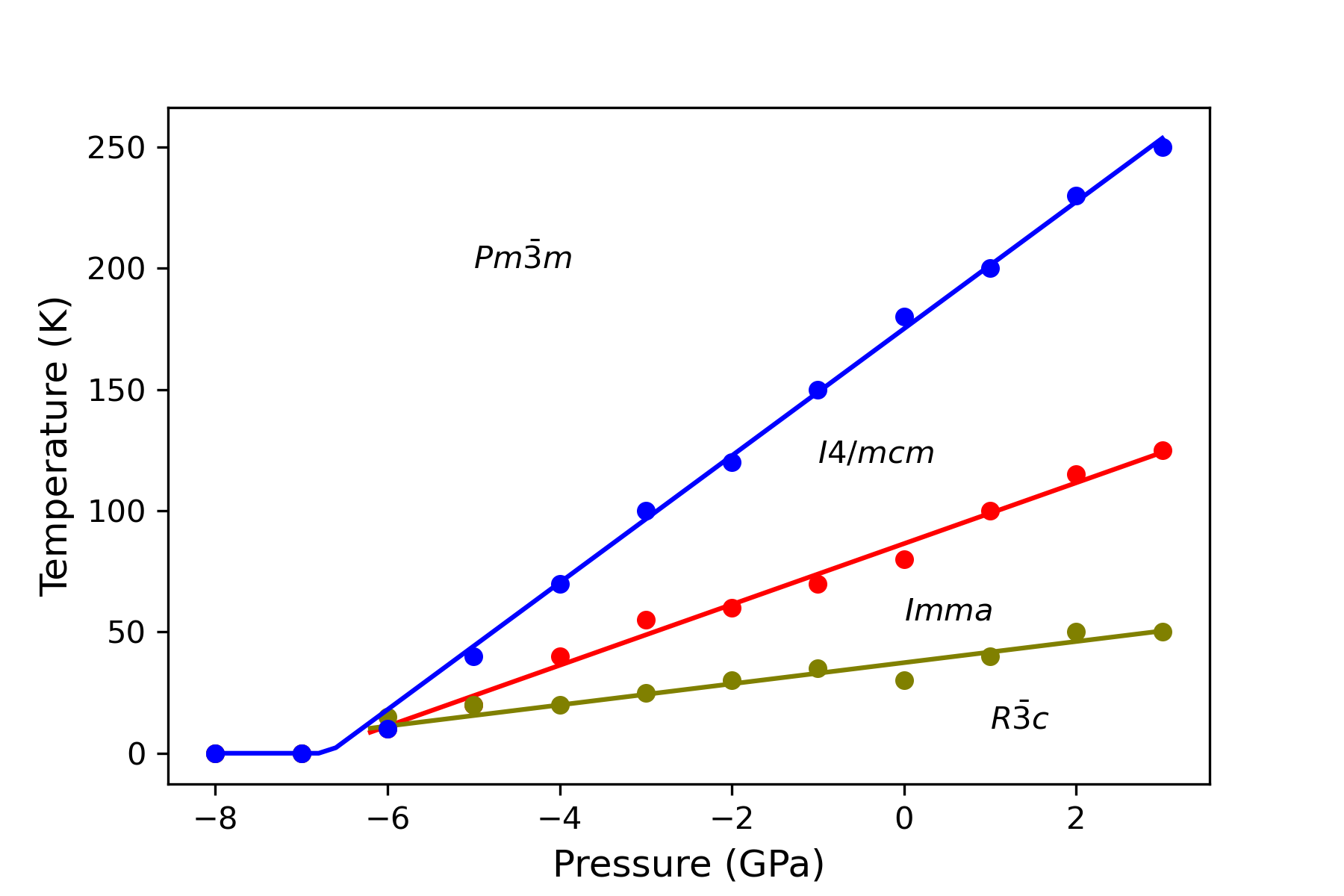}
\caption{\label{fig:TC-P}Temperature-pressure phase diagram of BaZrO$_3$ obtained with the shell model.}
\end{figure}

%%%%%%%%%%%%%%%%%%%%%%%%%%%%%%%%%%%%%%%%%%%%%%%%%%%%
The results of the simulations on the structural behavior obtained with the model are summarized in Figure~\ref{fig:TC-P} as the pressure dependence of the phase transition temperatures.
The values presented are those obtained directly from the simulations without applying pressure corrections or temperature rescaling.
As seen, the system undergoes a phase sequence with increasing temperature $R\bar{3}c \rightarrow Imma \rightarrow I4/mcm \rightarrow Pm\bar{3}m$, in which the AFD distortion around one pseudocubic axis successively disappear at each transition. 
This sequence emerges due to the nearly degenerate phases found at ambient pressure. However, the $I4/mcm \rightarrow Pm\bar{3}m$ transition is the unique observed experimentally when applying pressure to BaZrO$_3$, so this will be the only one that will be taken into account while the other two at lower temperatures will be neglected as they are considered an artifact of the model. 
As the negative pressure increases, the temperature range of the lower-symmetry phases diminishes,
noting in particular that the temperature-induced transition to the cubic phase is from the tetragonal for all pressures.
Our simulations predict that the pressure dependence of $T_C$ for that transition is essentially linear being
equivalent to the one observed experimentally in SrTiO$_3$~\cite{guen_10}.
The behavior of SrTiO$_3$ under pressure has been more studied in detail, and BaZrO$_3$ is expected to behave similarly.
The increment rate reported for SrTiO$_3$ is 18.5~K/GPa whereas the computed for BaZrO$_3$ is 25.7~K/GPa.
From this $T_C-P$ slope, we can estimate that the room temperature transition (300~K) would occur when applying $P$=5~GPa, which is lower than the value of 10~GPa recently reported from measurements at room temperature of pressure-induced $I4/mcm \rightarrow Pm\bar{3}m$ transition in BaZrO$_3$~\cite{toul_22}.
The difference would be related with the overestimation of the AFD instabilities in the model, causing the $T_C-P$ curve is shifted to lower pressures with respect to the still unknown experimental curve.
Applying a shift of +5~GPa will match the model value to that transition point, and the cubic structure in this case would still be unstable at $T$=0 and $P$=0. 
However, the cubic phase would be stable if we took as reference the value of 17~GPa from a past report by Yang~\cite{yang_14}.

\subsection{Varying the strength of the antiferrodistortive distortions}
\begin{figure}
\centering
\includegraphics[width=\linewidth]{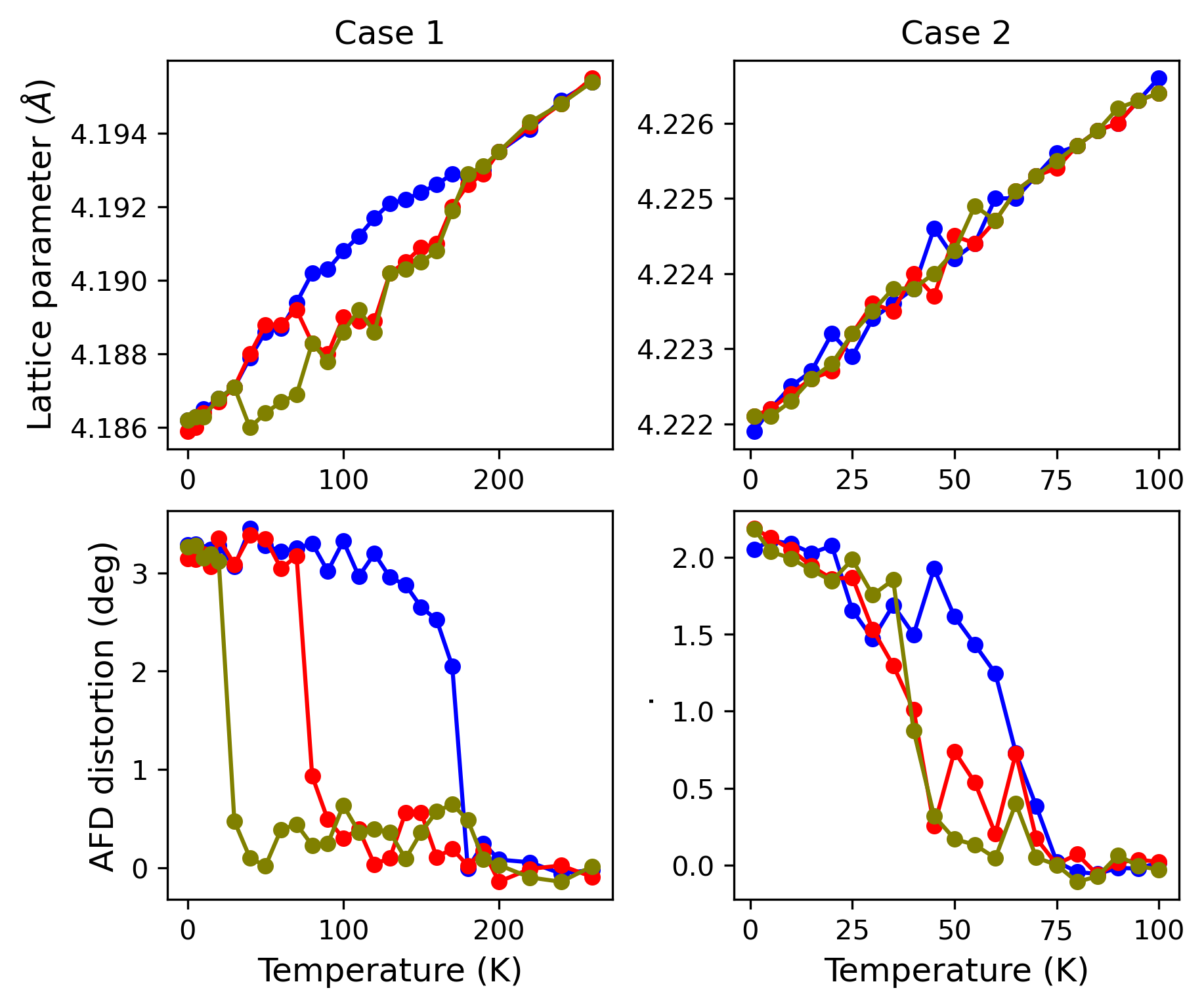}
\caption{\label{fig:Tem-e}Evolution with temperature of the three components of the lattice parameter (upper panels) and the antiferrodistortive rotation angle for the three directions of the rotational axis (bottom panels) for Cases 1 and 2.}
\end{figure}

From this point forward, we will concentrate on the behavior that BaZrO$_3$ could exhibit in various scenarios depending on the strength of the AFD distortions. To that end, we will look closely at three specific situations. The first one (Case 1) corresponds to a compound with well-marked AFD instabilities, that is, BaZrO$_3$ at $P$=0~GPa. In Case 2, the AFD distortions are slightly unstable ($P$=-4~GPa). Finally, Case 3 describes a system where the cubic phase is stable at all temperatures ($P$=-8~GPa).
Figure~\ref{fig:Tem-e} shows the temperature dependence of the pseudo-cubic lattice parameters and the AFD angles for cases 1 and 2. The results for Case 3 are not shown since the three lattice parameters have the same behavior and the parameter of rotational order is zero at all temperatures.

For pronounced AFD instabilities (Case 1), structural phases and transitions are clearly distinguished. In the low-temperature rhombohedral $R\bar{3}c$ phase, pseudo-cubic cell parameters are equal ($a = b = c$), and the rotation of the oxygen octahedra is along an [111] axis ($\theta^{x}=\theta^{y}=\theta^{z})$. 
The system becomes orthorhombic at T=40~K where $a<b=c$ and $\theta$ is along [011] and tetragonal at T=80~K (with $a=b<c$ and ($\theta^{x}=\theta^{y}=0,\theta^{z}\neq0)$; and finally cubic ($a=b=c$) at T=180~K where octahedral rotations in average disappear. 
When AFD instabilities are less pronounced (Case 2), it becomes challenging to differentiate the low symmetry phases.  
Lattice parameters do not show significant variations between them, although it is possible to identify the tetragonal phase through the rotational order parameter. 
Unlike Case 1, the $I4/mcm \to Pm\bar{3}m$ transition  shows a noticeable second order character, where the AFD order parameter $\theta$ continuously decreases until it vanishes at T$_C$.

\section{Atomic-scale properties}
\subsection{Radial distribution functions}

\begin{figure}
\centering
\includegraphics[width=\linewidth]{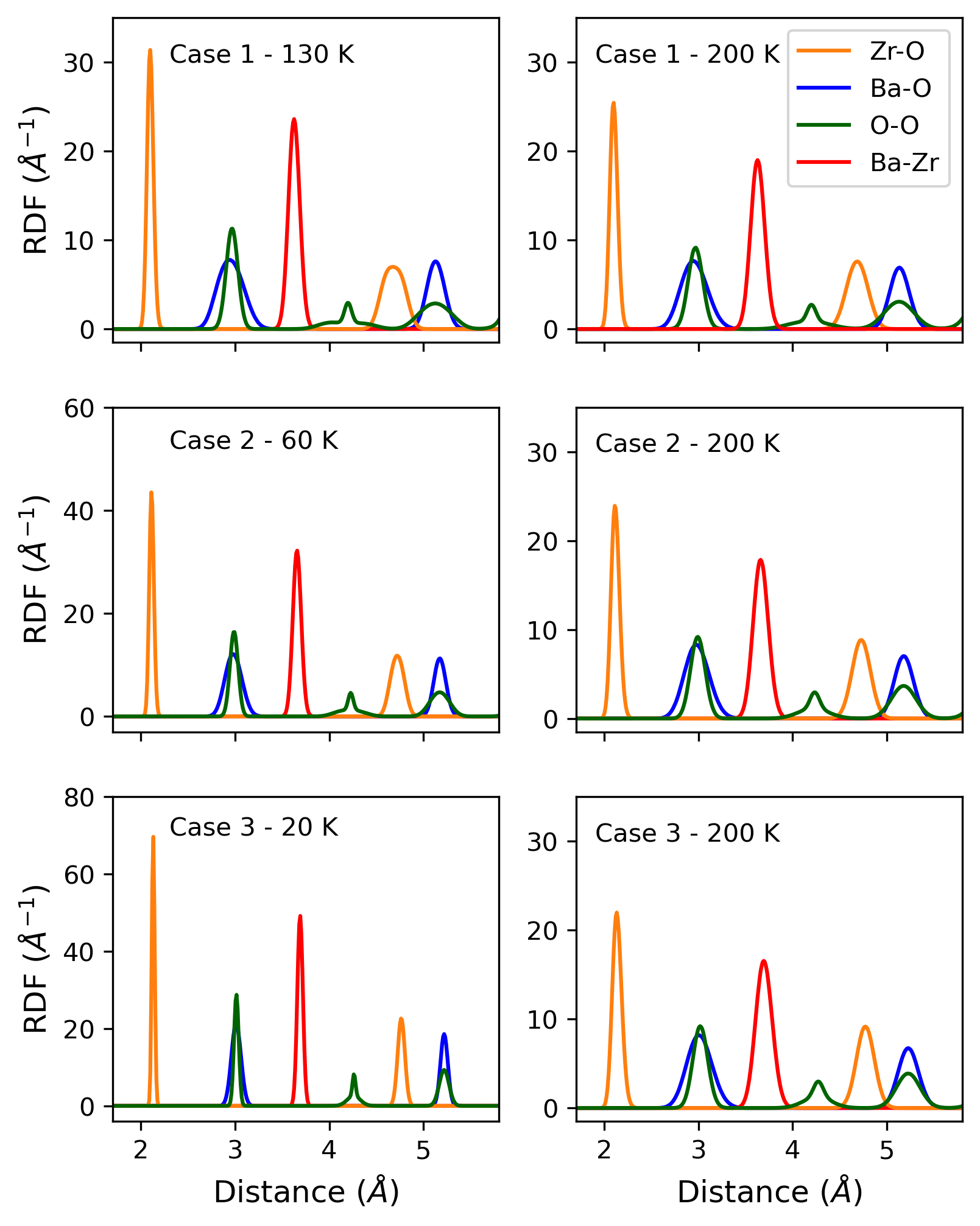}
\caption{\label{fig:RDF} Radial distribution functions obtained from the MD simulations for the three test cases at low temperature (130~K, 60~K, and 20~K) and 200~K shown in the left and right column, respectively. Note that the plots at 60~K and 20~K have different range for the RDF values.}
\end{figure}

To understand in more detail the microscopic characteristics of the analyzed cases, we inspect their radial distribution function (RDF), which is a scaled histogram of interatomic pair distance that provides precise information about the average local environment of each type of atom~\cite{terba_22}. The number of neighbors around a given atom for a coordination shell at a particular distance can be obtained by integrating RDF until that distance. Figure~\ref{fig:RDF} displays the RDFs for the three cases at low temperature and at 200~K, considering only the nearest pairs until a distance of 4.5~\AA. In each graph, the peaks observed from left to right correspond to the Zr-O (orange), Ba-O (blue), O-O (green) and Ba-Zr (red) pairs. Let us first note that the slight displacement in the peak positions among the three cases is due to their different lattice parameters. At lower temperature (left column), Case~1 and 2 present tetragonal symmetry while the system is cubic in Case~3,
but the differences between them are hard to distinguish due to the small structural distortions in the tetragonal phase plus the thermal effects.  
If the peaks could be distinguished from each other, in particular the twelve equivalent Ba-O bonds in the high symmetry phase should regroup in three peaks of four pairs each. 
This is precisely the reason why
in Case~1 the amplitude of the distribution Ba-O in the tetragonal phase at 130~K is slightly 
wider compared to that of the cubic phase at 200~K.
The distributions of the remaining pairs (Zr-O, O-O and Ba-Zr) look similar in all three cases providing interesting details of the low-temperature structures. First of all, the Ba-Zr pair distribution confirms that the relative position of both cations does not change when going from the low to the high symmetry structure. The sharp distribution of the Zr-O pair indicates that the Zr ions are equidistant from the nearest surrounding oxygen atoms and thus are located at the center of the O$_6$ cage. This distribution together with the corresponding to the O-O pair suggest that oxygen atoms form a rigid and undistorted octahedron. As temperature rises, the increased mobility of the ions due to thermal effects results in the broadening of the distributions, as seen at 200~K (right column). At this temperature, the three systems are in the same cubic phase, accordingly, no differences between the respective RDFs are appreciated. Note that the distributions of Ba-O and O-O reach a maximum at the same position and strikingly the former is wider than the latter, contrary to what happens in a conventional cubic structure. That is since oxygen is lighter than barium, the O-O pair distribution would be expected to exhibit larger thermal fluctuations opposite to the behavior found. Remarkably, a similar feature is also observed in the cubic structure of Case~3 at 20~K, where fluctuations are relatively low. These results can be rationalized in terms of coordinated displacements of the oxygen atoms, larger than those expected by thermal fluctuations. 
They agree with EXAFS measurements, and indeed it was from such “anomalous” behavior that the possibility of a low-temperature glassy phase was suggested~\cite{lebe_13}. Our simulations shed light on such anomalous behavior by showing that it is present in the cubic phase of BaZrO$_3$ regardless of whether or not a different phase exists at low temperature.

\begin{figure}
\centering
\includegraphics[width=\linewidth]{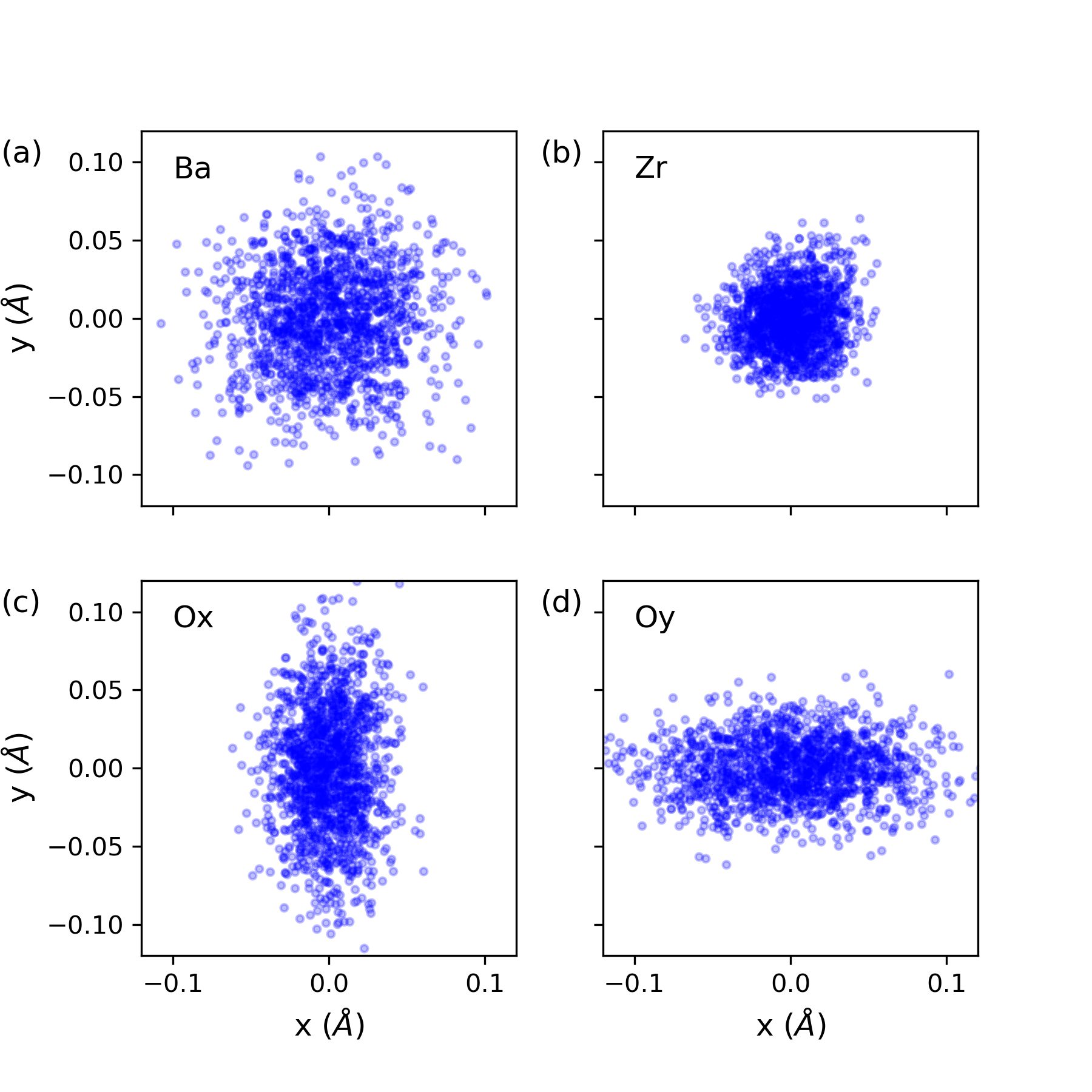}
\caption{\label{fig:sca} Ionic displacements with respect to their high-symmetry position in the cubic phase (origin) projected onto a $xy$ plane for different simulation times for BaZrO$_3$ under -8~GPa of pressure and at 20~K. (c) O$_x$ and (d) O$_y$ correspond the oxygen atoms that constitute the O$_6$ octahedron and are located in the direction of the respective Cartesian axes.}
\end{figure}

\subsection{Mean square displacements in the BaZrO$_3$ cubic phase}
To further clarify the microscopic behavior of the cubic phase, we plot the atomic positions with respect to the equilibrium ones as the simulation time elapses, and project them onto a particular $xy$ plane as shown in Figure~\ref{fig:sca}; similar graphics are obtained by analyzing projections on the other Cartesian planes $xz$ and $yz$. We focus on Case~3 as it exhibits a cubic structure all the way down to $T$=0~K without undergoing any phase transition. As it is observed, both cations Ba [Fig.~\ref{fig:sca}(a)] and Zr [Fig.~\ref{fig:sca}(b)] move isotropically around their ideal positions with the Ba displacements being larger than those of Zr, which is typical for 12-fold vs the 6-fold coordinated cations. The oxygen movement, instead, is clearly anisotropic with a narrow amplitude along the Zr-O-Zr bond and wider perpendicular to it, as seen for an oxygen aligned along the x-direction (O$_x$) in Fig.~\ref{fig:sca}(c) and the y-direction (O$_y$) in Fig.~\ref{fig:sca}(d) with respect to the Zr. Combining their movements on the $xy$ plane, together they would be compatible with the rotation of the octahedron around the $z$ axis. Similar results are found by observing the displacements of the other pairs of oxygen types (O$_x$-O$_z$ and O$_y$-O$_z$) in the corresponding planes. To quantify these displacements, we compute the mean square displacements $\left \langle u_{\alpha }  \right \rangle_{k}$ along the Cartesian direction $\alpha$ for each individual atom $k$ obtaining the following values: 
$\left \langle u_{1}  \right \rangle_{Ba}= 12\times 10^{-4} \AA^2$,
$\left \langle u_{1}  \right \rangle_{Zr}= 4\times 10^{-4} \AA^2$,
$\left \langle u_{1}  \right \rangle_{O_x}=18\times 10^{-4} \AA^2$,
$\left \langle u_{2}  \right \rangle_{O_x}=5\times 10^{-4} \AA^2$.
Our results at T = 20~K are underestimated compared with neutron diffraction data~\cite{perri_20} at T = 5~K, which give (18, 12, 58, 21)$\times10^{-4} \AA^2$. The discrepancies can be attributed to quantum-fluctuation effects not included in the present classical simulations, despite this, our model captures the ratio between the anisotropic displacements of oxygen atoms well.  Discrepancies tend to disappear when the comparison is made at higher temperatures. More importantly, our MD simulations support the presence of local atomic displacements compatible with rotations of the ZrO$_6$ octahedra in the cubic phase.

\subsection{Antiferrodistortive correlations in the BaZrO$_3$ cubic phase}

We now determine the rotation angles of the octahedra. Figures 7(a,b) display the time evolution of the rotation angle $\theta_{i}^{\beta }(t)$ around a Cartesian axis $\beta$ (z, in particular) for an arbitrary $i$-th cell belonging to the cubic phase of the Cases 1 and 3 at T = 200 K and 20 K, respectively. As seen, octahedra are effectively rotated with angles that fluctuate between  [-5$^{\circ}$,5$^{\circ}$] and [-2$^{\circ}$,2$^{\circ}$], but in both cases around an equilibrium value of zero. Similar results are obtained when inspecting the other rotational directions and different unit cells, and the same happens in all the cubic phases of three Cases analyzed. The question that now arises is whether octahedral rotations are correlated with those of neighboring cells. To answer it, we analyze their behavior in systems of $20\times 20\times20$ unit cells in size. The spatial correlation function defined as 
$\mathscr{C}_{\beta }^{\alpha }\left ( d \right )=\left \langle (-1)^{d}\theta{_{i}^{\alpha }}(t)\theta{_{i+d}^{\alpha }}(t) \right \rangle /\left \langle \theta_{i}^{\alpha } \right \rangle^{2}$ describes how two cells that are separated in the Cartesian direction $\beta$ by a number “$d$” of unit cells and have rotation axes along the cubic axis $\alpha$ are AFD coordinated. The brackets indicate spatial and temporal average over all $i$-th cells of the supercell and throughout the entire simulation time, respectively. 

\begin{figure}
\centering
\includegraphics[width=\linewidth]{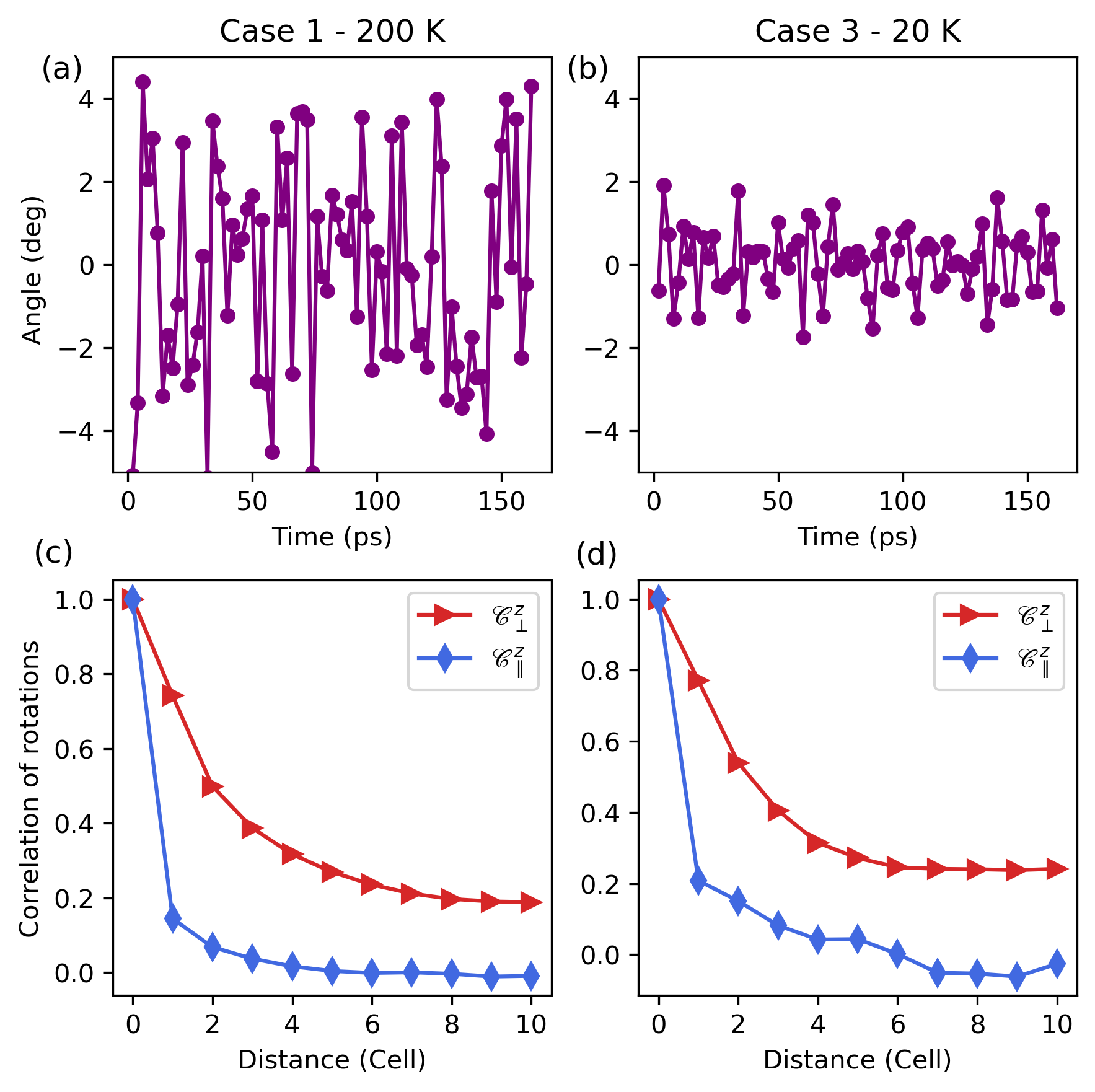}
\caption{\label{fig:Corr}(a,b) Time evolution of the rotation angle around the $z$ axis for an arbitrary cell and (c,d) spatial AFD correlation function between two cells separated by a number of unit cells 
along the perpendicular ($\mathscr{C}_{\bot}^{z}$) and parallel  ($\mathscr{C}_{\parallel}^{z}$) direction to the rotation axis $z$, belonging to the cubic phase of the Case~1 at 200~K (left panels) and Case~3 at 20~K (right panels).}
\end{figure}

At this point, it is convenient to remember that octahedra are hardly distorted according to the RDF information. So to maintain connectivity between neighboring octahedra in the plane normal to the rotation axis, the AFD (out-of-phase) distortions of almost rigid octahedra will form triangular wave-shaped patterns. Rotations in stacked layers perpendicular to the rotation axis, instead, are geometrically unconstrained and can present either in-phase or out-of-phase rotations~\cite{glazer_72}, although here we only inspect the AFD ones. Figures~\ref{fig:Corr}(c,d) compare the AFD correlations in the direction normal and parallel to the rotation axis. The correlation of the former is widely spread, going down to 20\% for a separation of approximately 7 unit cells, whereas the correlation in the parallel direction is low even for the nearest neighbors and continues to fall rapidly. Therefore, the apparently random oxygen displacements in the cubic phase are not, but also a remarkable AFD coordination exists among them on the plane normal to the rotation axis, which agrees with previous experimental and theoretical reports~\cite{lebe_13,perri_20,gran_20,levin_21}. 

\begin{figure*}
\centering
\includegraphics[scale=0.8]{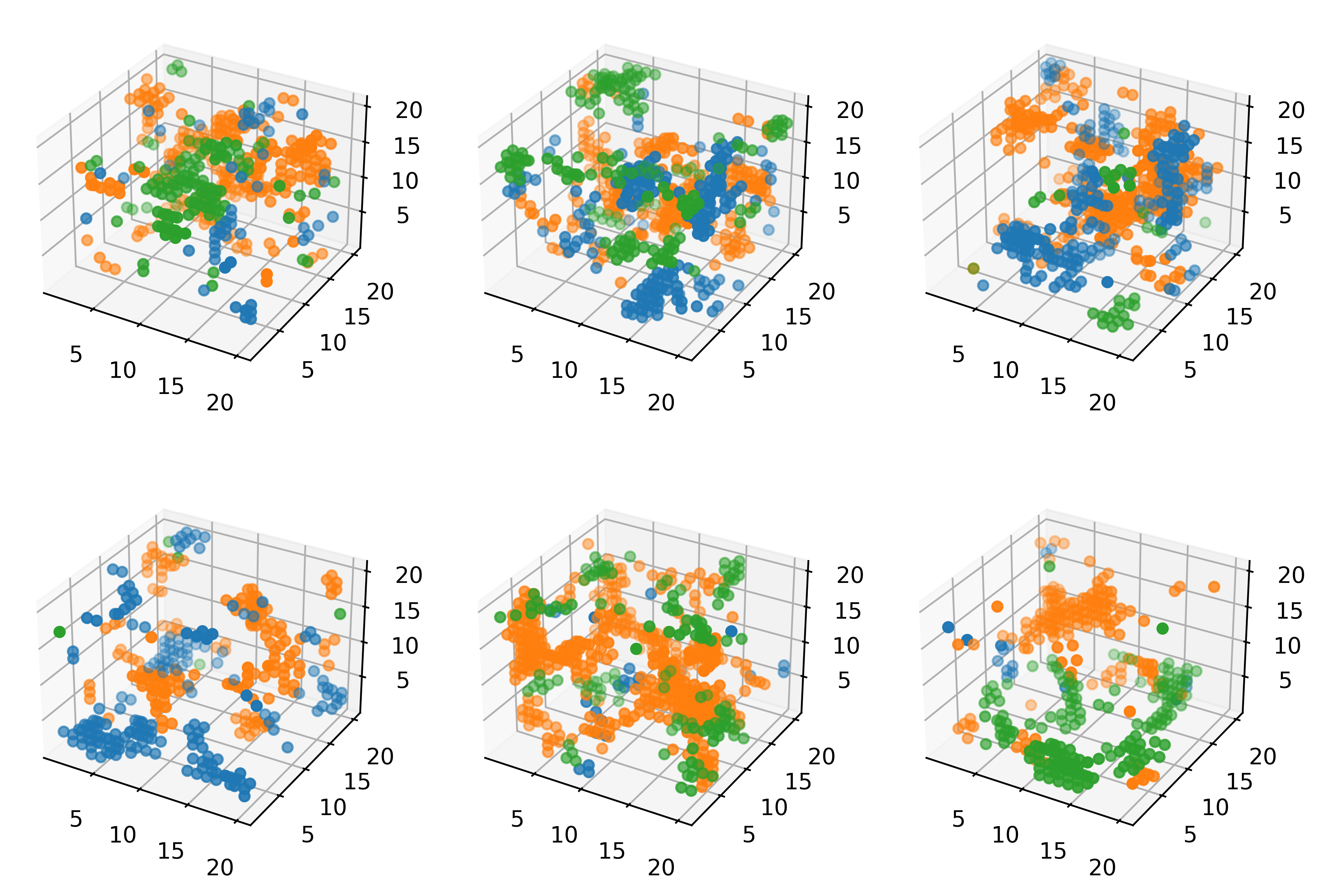}
\caption{\label{fig:snapshots}
Spatial distribution snapshots of unit cells with order parameters $\eta_x$, $\eta_y$ or $\eta_z$ equal to 2 (each identified with a different color), for BaZrO$_3$ in cubic phase for Case~1 at 180~K (upper panels) and Case~3 at 20~K (bottom panels) at three different simulation times.}
\end{figure*}

The observed correlations, however, are not enough to confirm the presence of relatively stable AFD domains within the cubic system. To investigate them, we evaluate the local order parameter $\eta$ for a given octahedron, presented in Ref.~\onlinecite{levin_21}. It is defined as $\eta = (\eta_{1}-\eta_{2})^{2} + (\eta_{4}-\eta_{3})^{2} $
where $\eta_{i}$ represents the fraction of octahedra in the $i$-th coordination shell that has the opposite sign of rotation relative to the central unit. Thus, $\eta$ measures the degree of AFD ordering of a cell with respect to the 32 nearest unit-cells that surround it (up to fourth neighbor shell). Values of $\eta$ = 0 and $\eta$ = 2 correspond to disorder and perfect order, respectively. Besides, the parameter for each unit cell has three values, one for each direction of the rotational axis. 
Indeed to find correlated structures, we consider only cells that have $\eta_x$, $\eta_y$ or $\eta_z$ equal to 2. As a result a cell could be ordered along one, two or three rotational axes with rotational patterns of the type $(0 0 a^-)$, $(0 a^-a^-)$, and $(a^-a^-a^-)$, respectively. Our results indicate that the probability of finding cells ordered along two or three axes simultaneously is negligible. We find instead unit-cell clusters ordered along a single Cartesian direction ($\eta_x$, $\eta_y$ or $\eta_z$) as shown in Figure~\ref{fig:snapshots} through snapshots taken at different simulation times. Each of these domains exhibits octahedral rotation patterns with $(0 0 a^-)$ type ordering, and as can be observed, they do not have a defined shape or size. Furthermore, using the advantage provided by MD simulations of monitoring the dynamic of the domains, we find they are not stable over time but are randomly created, evolve and disappear on any of the three possible rotation axes. These results completely agree with those obtained from the interpretation of electron diffraction and neutron scattering
measurments~\cite{levin_21}. But in addition, our simulations confirm that although the octahedral rotations in the cubic phase are significant, the correlated regions fail to form stable structures over finite time intervals. Supported by the three cases analyzed, we can further confirm that this happens in the cubic phase regardless of whether or not phase transitions associated with AFD distortions occur at lower temperatures, in other words, the AFD correlated regions exist without the need to be low-symmetry stable domains embedded in the cubic phase that could act as precursors of a phase transition. 

\section{Conclusions}

In this work, we have developed a shell model potential for BaZrO$_3$ by fitting {\it ab~initio} results of key properties. 
It is compatible with our previous model of BaTiO$_3$,
and it satisfactorily accounts for the change in the cubic lattice parameters, the loss of ferroelectricity, and the softening of the triply degenerate phonon mode at $R$-point of the cubic Brillouin zone experienced by BaZrO$_3$ with respect to BaTiO$_3$. 
This development allows the simulation not only of BaZrO$_3$ but also of
the Ba(Zr$_{x}$Ti$_{1-x}$)O$_3$ solid solution in the whole range of concentration without the necessity of any additional parameter.

To gain insights into the relationship between the experimentally observed structural anomalies in the cubic phase of BaZrO$_3$ 
and the possibility of an effective phase transition or the emergence of correlated domains,
we have examined three scenarios obtained by tuning the strength of the AFD instabilities.
By inspecting the cubic phase of the three cases at different temperatures
through a detailed atomic-scale analysis, we have found the following. The oxygen octahedra in BaZrO$_3$: (i) are barely distorted, with their Zr atom practically centered, (ii) present rotation angles that fluctuate with significant amplitudes,
(iii) are correlated with their closest neighbors in an AFD pattern along a single pseudocubic direction on the plane perpendicular to the rotation axis exhibiting ($00a^-$)-type order, (iv) form nanoregions antiferrodistortively coordinated that exhibit a dynamical and unstable nature. 
In conclusion, our results support the fact 
that the experimentally observed structural anomalies in cubic BaZrO$_3$ can exist regardless of whether or not structural phase transitions related to antiferrodistortive distortions occur at lower temperatures.

\hspace{0.5cm}
\begin{acknowledgments}
We acknowledge the computing time at the CCT-Rosario Computational Center. This work was sponsored by the National
Scientific and Technical Research Council (Consejo Nacional de Investigaciones Científicas y Técnicas, CONICET) (PUE-IFIR 2017-2022 No.~0076 and PUE-IFIS 2016-2022 No.~0054), the National University of Rosario (Universidad Nacional de Rosario, UNR) (No.~80020180300068UR) and the  National University of Litoral (Universidad Nacional del Litoral) (CAI+D2020 No.~50620190100068LI). M.G.S. acknowledges support from the National University of Rosario Research Council (Consejo de Investigaciones de la Universidad Nacional de Rosario, CIUNR).
\end{acknowledgments}

\bibliography{bzo}

\end{document}